\documentclass[conference,a4paper]{IEEEtran}
\usepackage{xcolor}
\usepackage{balance}

\ifCLASSINFOpdf
  \usepackage[pdftex]{graphicx}
\else
  \usepackage[dvips]{graphicx}
\fi
\usepackage{amsmath,amssymb}
\ifCLASSOPTIONcompsoc
 \usepackage[caption=false,font=normalsize,labelfont=sf,textfont=sf]{subfig}
\else
 \usepackage[caption=false,font=footnotesize]{subfig}
\fi

\usepackage{bm}
\newcommand{\R}{\mathbb{R}}
\newcommand{\C}{\mathbb{C}}

\newcommand{\ie}{\textit{i.e.}\/, }
\newcommand{\eg}{\textit{e.g.}\/, }
\newcommand{\cf}{\textit{cf.}\/, }

\providecommand*{\mrm}[1]{\mathrm{#1}}
\providecommand*{\unit}[1]{\ensuremath{\mrm{\,#1}}}
\providecommand*{\eu}{\ensuremath{\mrm{e}}}
\providecommand*{\iu}{\ensuremath{\mrm{i}}}

\providecommand*{\diff}{\operatorname{d}\!}
\renewcommand{\Im}{\ensuremath{\mrm{Im}}}	

\newtheorem{theorem}{Theorem}[section]

\begin{document}
%
\title{Uniform error bounds for fast calculation of approximate Voigt profiles}

\author{\IEEEauthorblockN{
Sven~Nordebo\IEEEauthorrefmark{1},   
}                                     
\IEEEauthorblockA{\IEEEauthorrefmark{1}
Department of Physics and Electrical Engineering, Linn\ae us University,   351 95 V\"{a}xj\"{o}, Sweden. E-mail: sven.nordebo@lnu.se} 
}



\maketitle

\begin{abstract}
The broadband line-by-line analysis of radiative transfer in the atmosphere is extremely demanding with regard to computational resources.
As a remedy, we present here the calculation of uniform error bounds for approximating the classical Voigt profile. 
A new ``full'' Voigt profile, which can be expressed as a combination of two Faddeeva evaluations, is also presented and included in the analysis. 
The uniform bounds can be used to rigorously determine the domains on which
the Voigt profiles can be approximated by the corresponding Lorentz profiles to any desired accuracy. The bounds can furthermore be employed
to make a fast and efficient estimate of the most significant lines to be included in a subband adaptive line selection strategy.
By using a realistic numerical example of radiative transfer in the atmosphere, we demonstrate that these approximation approaches
are able to reduce the computational time of the associated line-by-line analysis by several orders of magnitude with little loss of accuracy.
\end{abstract}

\vskip0.5\baselineskip

\section{Introduction}

Even though there are advanced theories of molecular line shapes taking
line mixing and velocity changes into account  \cite{Filippov+etal2002,Filippov+Tonkov1993,Gordon1966,Gordon1967,Hartmann+etal2008,Ngo+etal2013,Rosenkranz1975}, 
most databases and radiative transfer codes are still using the classical Voigt profile, 
see \eg \cite{Berk+Hawes2017,Gordon+etal2017b,Rothman+etal1998,Tennyson+etal2014}.
A simple explanation for this is probably the fact that
the broadband line-by-line analysis of radiative transfer in the atmosphere is extremely demanding with regard to computational resources.
Quoting Liou \cite[p.~126]{Liou2002}: ``The computer time required for line-by-line calculations, even with the availability of a supercomputer, is formidable. 
This is especially true for flux calculations in which an integration over all absorption bands is necessary.''
Typically, hundreds of thousands of spectral lines must be resolved to a resolution corresponding to millions of
frequency points at each atmospheric level. And at the same time, most of these calculations could have been performed
based on the simple Lorentzian profile rather than by using the more complex Voigt profile, and most of the spectral line contributions could have
been disregarded with very little impact on the final accuracy of the computation.
Much of these problems have been overcome with recent line-by-line algorithms based on simplifications, approximations and pre-computed line data,
see \eg \cite{Berk+Hawes2017}. As an aid in the future development, improvement and extension of similar algorithms,
this paper is providing rigorous bounds for approximating the Voigt profile by the much simpler Lorentz profile.

The Voigt profile is the convolution between the Lorentzian profile and the Gaussian.
And even though there exists computer codes for efficient calculation of the associated Faddeeva function, see \eg \cite{Abrarov2020,Abrarov+Quine2011,Schreier1992,Tennyson+etal2014}, 
it is much faster just to compute the corresponding Lorentz profile. 
Moreover, the superexponential Gaussian profile is a highly localized function, a property that can be exploited to significantly 
accelerate the calculation of an approximate Voigt profile. 
In this paper we formulate and prove a theorem which 
quantifies the fact that the Voigt profile converges uniformly towards the Lorentzian profile in two different regards: 
(1) with respect to the Lorentzian half-width normalized by the Gaussian half-width, 
and where the approximation error is measured over the whole frequency axis, and
(2) with respect to the line width normalized by the Gaussian half-width, and beyond which the far wing approximation error is measured.
Thus, (1) if the Lorentzian is sufficiently broad in comparison to the Gaussian, it approximates the Voigt profile everywhere.
And (2) in the far wings, the Voigt profile can always be approximated by the Lorentzian, provided that the beginning of the far wing is suitably defined.
In this paper we will show how the Voigt (or Faddeeva) function can be used to pre-calculate the associated threshold values
in order to achieve any desired approximation accuracy.
The related computations based on thresholding and sorting can readily and efficiently be implemented in a computer code. 
In a typical application of radiative transfer in the atmosphere, most of the frequency
points to be evaluated will fall into the category where these approximations are applicable, and hence the potential to save computation time is
almost the same as replacing the Voigt computation for the Lorentzian.

However, the line-by-line computation of broadband radiative transfer in the atmosphere is still huge.
To further reduce the computational complexity, a subband (or block) adaptive line selection procedure is presented here. 
For each spectral line exterior to a fixed and relatively narrow subband, the new error bounds can be used to
determine precisely where the Lorentz approximation is applicable, and then very fast and effectively
estimate its in-band contribution to any desired accuracy. Thus, the ``exterior'' spectral lines that are estimated to have a negligible
in-band contribution can be excluded from the computation. In-band lines are always included, 
and the computation can then proceed with the next subband.
Even though this is a suboptimal approach, it is a simple, pragmatic and very effective way to include only the lines
that are the most significant, and the procedure can be tuned with very few threshold parameters to control the accuracy.
By using realistic numerical examples of radiative transfer in the atmosphere,
we will demonstrate that the combination of the two approximation approaches described above can
reduce the computational time of the line-by-line analysis by several orders of magnitude with very little loss of accuracy.

In this work we do not attempt to compare the performance of our algorithm with other highly sophisticated commercial 
codes for radiative transfer such as the MODTRAN 6 line-by-line algorithm reported in \cite{Berk+Hawes2017}.
However, a few general remarks can be made as follows. 
First, the purpose of our numerical example has been to evaluate and compare the proposed
approximation methods by using an algorithm that is as simple and straightforward and possible.
A layer-recursive algorithm for radiative transfer in a one-dimensional plane-parallel and non-scattering 
atmosphere has therefore been developed for this purpose. 
The frequency resolution has been chosen based on the (frequency dependent) Doppler broadening in the 
upper atmosphere at $65\unit{km}$ height, and is therefore extremely dense with 
a resolution starting at $75\cdot 10^{-6}$\unit{cm^{-1}} in the lower $100\unit{cm^{-1}}$ band.
All profiles are calculated on-the-fly, based on line parameters that
have been retrieved from the HITRAN database \cite{Gordon+etal2017b,Rothman+etal1998}.
Similar as with the MODTRAN algorithm, our algorithm is computing Voigt profiles with their line centers within certain narrow subbands.
The main difference is the way in which the exterior tail contributions are selected and computed. Our algorithm
is employing the adaptive line selection procedure as described above, whereas the MODTRAN algorithm is employing
a fixed 25\unit{cm^{-1}} maximal distance for their line selection \cite[p.~544--545]{Berk+Hawes2017}. 
The MODTRAN algorithm is furthermore employing pre-computed line shape data which is stored on a fine 
temperature and coarse pressure grid, and which can then readily be scaled for the adequate pressure on-the-fly.
The latter procedure, which is based on Pad\'{e} approximations of Voigt sums, is no doubt an effective way of making the algorithm very fast.

We will also introduce and incorporate in the analysis the ``full'' Voigt profile.
The full Voigt profile is based on the convolution between the Gaussian profile and the 
``full'' Lorentzian profile which is just the classical Lorentz model without performing the traditional resonance approximation.
It will be shown that the full Voigt profile can be computed based on two Faddeeva calculations, instead of just one as with the 
traditional Voigt profile. However, since most of the computations in a typical application can be performed under the
full Lorentz approximation in very much the same way as have been described above, the approximation of the full Voigt profile 
is only slightly more complex in comparison to the approximation of the traditional Voigt profile.

The rest of the paper is organized as follows. In Section \ref{sect:jqsrt_Lorprofiles} is given a brief account on the ``full'' Lorentz profile,
and its uniform convergence properties with respect to the traditional resonance approximation. 
In Section \ref{sect:jqsrt_voigtprofiles} is given the corresponding Voigt profiles, and in particular the ``full'' Voigt profile which can be expressed as a combination of
two Faddeeva calculations. The details of the derivation are given in the Appendix \ref{sect:voigt_FullVoigtderivation}. 
In Section \ref{sect:jqsrt_uniformapprox} is given the theorem which quantifies the
uniform approximation properties of the Voigt profiles with regard to the corresponding Lorentz profiles.
The details of the proof are given in the Appendix \ref{sect:voigt_uniform}.
The subband adaptive line selection strategy is described in Section \ref{sect:LS} and the numerical examples are given in Section \ref{sect:numex}.
The paper is concluded with a short summary and a discussion.

\section{The classical Lorentz profile}\label{sect:jqsrt_Lorprofiles}
In a nonscattering atmosphere in the long wave regime, the absorption cross section $k_\nu$ 
of a molecule is the same as its extinction (or total) cross section $\sigma_\mrm{t}$.
Based on the classical optical theorem \cite[pp.~227--230, 272]{Kristensson2016}, \cite[pp.~18-20]{Newton1982} 
the absorption (extinction) cross section $k_\nu$ can then be modeled as
\begin{equation}\label{eq:voigt_knudef}
k_\nu=\Im \{2\pi\nu\alpha\},
\end{equation}
where $\alpha$ is the polarizability, $\nu=\lambda^{-1}$ is the wavenumber and $\lambda$ the wavelength in vacuum.
This means that we can obtain the absorption coefficient as $k_\nu=\Im\{h(\nu)\}$ where $h(\nu)$ is a 
Herglotz function\footnote{A Herglotz function $h(z)$ is a holomorphic function 
defined on the open upper half-plane $\C^+=\{z\in\C|\Im\{z\}>0\}$ and where its imaginary part is non-negative, \ie $\Im\{h(z)\}\geq 0$ for $z\in\C^+$.} 
in the complex variable $\nu$, and remember that it is really the extinction coefficient that we are modeling.
Basic properties of Herglotz functions can be found in \eg \cite{Akhiezer1965,Gesztesy+Tsekanovskii2000,Kac+Krein1974,Nedic+etal2019,Nussenzveig1972}.
Perhaps the most simple, and the most widely used model of the molecular polarizability is the classical Lorentz model
\begin{equation}\label{eq:voigt_alphadef}
\alpha=S\frac{1}{\pi^2}\frac{1}{\nu_0^2-\nu^2-\iu 2\gamma\nu},
\end{equation}
where $S$ is the line strength, $\nu_0$ the transition frequency (including the pressure shift) and $\gamma$ the Half-Width-Half-Maximum (HWHM) parameter
modeling the collision induced line broadening. Notably, the expression \eqref{eq:voigt_alphadef} can be derived by using 
a simple classical phenomenological model, as well as by using the more comprehensive perturbation techniques of
quantum mechanics, see \eg \cite[pp.~232-233]{Bohren+Huffman1983}, \cite{Cai+etal1987}, \cite[pp.~117]{Grynberg+etal2010}, 
\cite[pp.~104-108]{Schatz+Ratner2002} and \cite[pp.~1808-1809]{Sipe+Kranendonk1974}.

The Herglotz function modeling the line shapes in \eqref{eq:voigt_knudef} is now given by
\begin{equation}\label{eq:voigt_hnudef}
h(\nu)= \frac{2}{\pi}\frac{\nu}{\nu_0^2-\nu^2-\iu 2\gamma\nu},
\end{equation}
and its imaginary part is
\begin{equation}\label{eq:voigt_FLdef}
f_\mrm{FL}(\nu)=\Im\{h(\nu)\}=\frac{4}{\pi}\frac{\gamma \nu^2}{\left(\nu_0^2-\nu^2\right)^2+4\gamma^2\nu^2}.
\end{equation}
Here, we will refer to \eqref{eq:voigt_FLdef} as the ``full Lorentz'' profile to distinguish it from the usual resonance approximation given below.
The asymptotic behavior of \eqref{eq:voigt_hnudef} is readily found as
\begin{equation}\label{eq:voigt_hLFHFassymptotic}
h(\nu)=\left\{\begin{array}{l}
a_1\nu+o(\nu)  \quad \textrm{as}\ \nu\rightarrow 0, \vspace{0.2cm}\\
b_{-1}\nu^{-1}+o(\nu^{-1}) \quad \textrm{as}\ \nu \rightarrow \infty,
\end{array}\right.
\end{equation}
where $o(\cdot)$ denotes the small ordo \cite[p.~4]{Olver1997} and 
where $a_1=2/\pi\nu_0^{2}$ and $b_{-1}=-2/\pi$. 
The Herglotz function in \eqref{eq:voigt_hnudef} is symmetric, and the following 
two sum rules \cite{Nedic+etal2019} apply
\begin{equation}\label{eq:voigt_sum1}
\frac{2}{\pi}\int_{0}^\infty \Im\{h(\nu)\}\diff\nu=-b_{-1}=\frac{2}{\pi},
\end{equation}
and
\begin{equation}\label{eq:voigt_sum2}
\frac{2}{\pi}\int_{0}^\infty \frac{\Im\{h(\nu)\}}{\nu^2}\diff\nu=a_{1}=\frac{2}{\pi\nu_0^2}.
\end{equation}
The first sum rule in \eqref{eq:voigt_sum1} establishes that the full Lorentz profile is normalized over $\R^+=[0,\infty)$, and the second
gives a sum rule relating the extinction cross section $\sigma_\mrm{t}$
to the static polarizability of the line, $\int_0^\infty\diff\nu \sigma_\mrm{t}/\nu^2=\pi^2 \alpha(0)=S/\nu_0^2$, 
\cf \cite{Cai+etal1987,Sohl+etal2007a,Bernland+etal2011b,Nedic+etal2019}.

The classical Lorentz profile used in most radiative transfer analysis is given by
\begin{equation}\label{eq:voigt_Ldef}
f_\mrm{L}(\nu-\nu_0)=\frac{1}{\pi}\frac{\gamma}{(\nu-\nu_0)^2+\gamma^2},
\end{equation}
see \eg \cite[p.~73]{Bohren+Clothiaux2006}, \cite[p.~21-23]{Liou2002}, \cite[pp.~263-266]{Petty2006} and \cite[Eq.~(A14)]{Rothman+etal1998}.
The classical Lorentz profile above can readily be obtained as an approximation of \eqref{eq:voigt_FLdef} valid for $\nu$ close to resonance at $\nu_0$.
It is emphasized, however, that \eqref{eq:voigt_Ldef} can also be derived from first principles, as in \cite[pp.~77-78]{Hartmann+etal2008}.
It is noticed that the convergence of the aforementioned approximation is somewhat subtle, in particular for small $\nu$ and small $\gamma$
since $f_\mrm{FL}(0)=0$ and $f_\mrm{L}(-\nu_0)=\gamma/\pi(\nu_0^2+\gamma^2)$. To analyse this situation
we consider the relative error
\begin{multline}
E^\mrm{FL}_{\gamma,\nu_0}(\nu)=\frac{f_\mrm{FL}(\nu)-f_\mrm{L}(\nu-\nu_0)}{f_\mrm{L}(\nu-\nu_0)} \\
=\frac{(\nu-\nu_0)^3(3\nu+\nu_0)}{(\nu_0+\nu)^2(\nu_0-\nu)^2+4\gamma^2\nu^2},
\end{multline}
yielding the upper bound
\begin{equation}
\left| E^\mrm{FL}_{\gamma,\nu_0}(\nu) \right|\leq \frac{|\nu-\nu_0|(3\nu+\nu_0)}{(\nu_0+\nu)^2}.
\end{equation}
By assuming that $|\nu-\nu_0|\leq B$, it is readily seen that
\begin{equation}\label{eq:voigt_uniformFL1}
\left| E^\mrm{FL}_{\gamma,\nu_0}(\nu) \right|\leq \frac{B\nu}{\nu_0^2}\leq \frac{B(\nu_0+B)}{\nu_0^2} \quad \mrm{for} \quad \nu>\nu_0>0,
\end{equation}
and where $\nu<\nu_0+B$, and
\begin{equation}\label{eq:voigt_uniformFL2}
\left| E^\mrm{FL}_{\gamma,\nu_0}(\nu) \right|\leq \frac{B\nu_0}{\nu^2}\leq \frac{B(\nu+B)}{\nu^2} \quad \mrm{for} \quad 0<\nu<\nu_0,
\end{equation}
and where $\nu_0<\nu+B$.  For a fixed bandwidth $B_\gamma$ (related to the half-width $\gamma$) and $|\nu-\nu_0|\leq B_\gamma$, 
it is now readily seen that $E^\mrm{FL}_{\gamma,\nu_0}(\nu)$ converges uniformly to zero as $\nu_0\rightarrow\infty$.
It is emphasized that this result is merely a mathematical property of the line shapes which we will need later, 
and we do not intend to let the limit of large transition frequencies $\nu_0$ violate the assumption about a non-scattering atmosphere.

For most practical purposes the analytically more tractable Lorentz profile \eqref{eq:voigt_Ldef} is an extremely good approximation   
of \eqref{eq:voigt_FLdef}, except for some far wing calculations in the infrared windows where the density of spectral lines are very sparse.
It may also be of interest to observe that the classical Lorentz profile \eqref{eq:voigt_Ldef} is not symmetric, and
hence does not correspond to a symmetric Herglotz function as would be physically expected in relation to the optical theorem \eqref{eq:voigt_knudef}, 
\cf \cite{Bernland+etal2011b,Nedic+etal2019,Sohl+etal2007a}.
In particular, since $f_\mrm{L}(-\nu_0)\neq 0$ there is no sum rule in the form of \eqref{eq:voigt_sum2} relating the $\nu^{-2}$ moment of $f_\mrm{L}(\nu-\nu_0)$ 
to the static polarizability of the line.

\section{Voigt profiles}\label{sect:jqsrt_voigtprofiles}

The classical Voigt profle is defined by the convolution integral
\begin{equation}\label{eq:voigt_fVdef}
f_\mrm{V}(\nu)=\int_{-\infty}^{\infty}f_\mrm{G}(t)f_\mrm{L}(\nu-t)\mrm{d}t,  \quad \nu\in\R,
\end{equation}
where $f_\mrm{G}(\nu)$ is the Gaussian profile 
\begin{equation}\label{eq:voigt_fGdef}
f_\mrm{G}(\nu)=\sqrt{\frac{\ln 2}{\pi}}\frac{1}{\alpha}\eu^{-\nu^2\ln 2/\alpha^2}, \quad \nu\in\R,
\end{equation}
modeling the Doppler broadening and where $\alpha$ is the corresponding HWHM parameter.
Notice that the absorption line associated with the line center frequency $\nu_0$ is 
given by the translation $f_\mrm{V}(\nu-\nu_0)$.

It is well known that the Voigt profile \eqref{eq:voigt_fVdef} can be computed by means of the Faddeeva function
\begin{equation}\label{eq:voigt_wdef}
w(z)=\frac{\iu}{\pi}\int_{-\infty}^\infty\frac{\eu^{-t^2}}{z-t}\mrm{d}t,
\end{equation}
for which there exists many efficient numerical codes, see \eg \cite{Abrarov2020,Abrarov+Quine2011,Schreier1992,Tennyson+etal2014}. 
In particular, by extending the integrand above with the complex conjugate of its denominator,  it can be readily shown that
\begin{equation} \label{eq:voigt_fVjresult}
f_\mrm{V}(\nu)=
\sqrt{\frac{\ln 2}{\pi}}\frac{1}{\alpha}\Im\left\{ \iu w(x+\iu y)\right\},
\end{equation}
where $x=\nu\sqrt{\ln 2}/\alpha$ and $y=\gamma\sqrt{\ln 2}/\alpha$. It may be noticed that $\iu w(z)$ is a symmetric
Herglotz function generated by the Gaussian density, whereas $f_\mrm{V}(\nu-\nu_0)$ is the density of a non-symmetric (shifted) Herglotz function.

Now, we define also the ``full Voigt'' profile by
\begin{equation}\label{eq:voigt_fFVdef}
f_\mrm{FV}(\nu)=\int_{-\infty}^{\infty}f_\mrm{G}(t)f_\mrm{FL}(\nu-t)\mrm{d}t,  \quad \nu\in\R,
\end{equation}
where $f_\mrm{FL}(\nu)$ is the full Lorentz profile given by \eqref{eq:voigt_FLdef}.
In Appendix \ref{sect:voigt_FullVoigtderivation} is shown that the full Voigt profile can be computed by means of two Faddeeva calculations as
$f_\mrm{FV}(\nu)=\Im\{h_\mrm{FV}(\nu)\}$ where
\begin{multline}\label{eq:voigt_fFVjresult}
h_\mrm{FV}(\nu)=\sqrt{\frac{\ln 2}{\pi}}\frac{1}{\alpha}\left(
\left(-\frac{\gamma}{a}+\iu \right)w\left((\nu+a+\iu\gamma)\frac{\sqrt{\ln 2}}{\alpha}\right) \right. \\
\left. +\left(\frac{\gamma}{a}+\iu \right)w\left((\nu-a+\iu\gamma)\frac{\sqrt{\ln 2}}{\alpha}\right)
\right),
\end{multline}
and where $a=\sqrt{\nu_0^2-\gamma^2}$ and $\nu_0>\gamma$.


\section{Uniform approximation of Voigt profiles}\label{sect:jqsrt_uniformapprox}

\subsection{The classical Voigt profile}
We define the relative error between the classical Voigt and Lorentz profiles as
\begin{equation}\label{eq:voigt_Ealphagammadef}
E^\mrm{V}_{\alpha,\gamma}(\nu)=\frac{f_\mrm{V}(\nu)-f_\mrm{L}(\nu)}{f_\mrm{L}(\nu)}, \quad \nu\in\R,
\end{equation}
and which can be manipulated to read
\begin{multline}\label{eq:voigt_Ealphagammaexpr}
E^\mrm{V}_{\alpha,\gamma}(\nu)
=\sqrt{\frac{\ln 2}{\pi}}\frac{1}{\alpha}\int_{-\infty}^{\infty}\eu^{-t^2\ln 2/\alpha^2}\frac{t(2\nu-t)}{\gamma^2+(\nu-t)^2}\mrm{d}t,
\end{multline}
which is an even function of $\nu$.
It is readily seen that $E^\mrm{V}_{\alpha,\gamma}(\nu)={\cal O}\{\nu^{-1}\}$ for fixed values of $\alpha$ and $\gamma$ and
where ${\cal O}\{\cdot\}$ denotes the big ordo \cite[p.~4]{Olver1997}. Hence, for given values of $\alpha$ and $\gamma$
the absolute error $|E^\mrm{V}_{\alpha,\gamma}(\nu)|$ will be arbitrarily small for sufficiently large $\nu$.
The superexponential Gaussian profile is furthermore highly localized and converges to the Dirac delta function as $\alpha\rightarrow 0$.
Hence, we certainly have $\lim_{\alpha\rightarrow 0}f_\mrm{V}(\nu)=f_\mrm{L}(\nu)$ when $\gamma$ and $\nu$ are fixed.
However, as shown in the Appendix~\ref{sect:voigt_uniform}, the convergence is in fact uniform over $\nu\in\R$ as $\alpha\rightarrow 0$ and $\gamma$ is fixed.
From the properties mentioned above it is now very close at hand to formulate the following theorem providing two simple statements, or criteria, 
for achieving a prescribed uniform error tolerance. 

\begin{theorem}\label{thm:Voigtaccuracy}
For any $\epsilon>0$ and $n_1>0$ there exists positive real numbers $n_2$ and $n_3$ 
such that the following two statements regarding the line parameters $(\alpha,\gamma)$ hold
\begin{equation}\label{eq:voigt_Accuracycriterion}
\left\{\begin{array}{l}
 \gamma/\alpha>n_2  \Rightarrow  \max_{\nu\in \R} |E^\mrm{V}_{\alpha,\gamma}(\nu)|<\varepsilon \vspace{0.2cm} \\ 
 \gamma/\alpha>n_1 \Rightarrow  \max_{|\nu|>n_3\alpha} |E^\mrm{V}_{\alpha,\gamma}(\nu)|<\varepsilon.
\end{array}\right.
\end{equation}
\end{theorem}
The lower limit relating to $\gamma/\alpha>n_1$ above is probably not needed as $n_1$ can be chosen to be arbitrarily small,
but is included here with the purpose of simplifying the proof of the theorem. 
In addition, in a practical application the parameter $\gamma/\alpha$ is always bounded from below by a nonzero number $n_1$.
The Theorem \ref{thm:Voigtaccuracy} will be proved below, but let us first give a few comments on its application.

For a given set of line parameters $(\alpha,\gamma)$ the first statement in \eqref{eq:voigt_Accuracycriterion} is the stronger one,
and it can be used as a criterion to determine whether the Voigt profile can be approximated by the Lorentz profile within the given error tolerance $\varepsilon$
on the whole of the frequency axis $\nu\in\R$.
If the condition in the first statement is not satisfied, then the second (weaker) statement asserts that it is sufficient to calculate the Voigt profile within
the (usually very small) limited frequency range $|\nu|\leq n_3\alpha$, and hence that the Lorentz profile can be used for $|\nu|> n_3\alpha$. 
The practical usefulness of the theorem stems from the fact that it is numerically much more efficient
to calculate the Lorentz profile in comparison to the Voigt profile, and that simple comparative conditions such as  $|\nu|\leq n_3\alpha$ above
can readily be implemented in software by using sorting routines at very low computational cost.
The application of Theorem \ref{thm:Voigtaccuracy} can now be summarized as follows.
Assume that $\varepsilon>0$ and $\gamma/\alpha>n_1$. Then there are positive real numbers $n_2$ and $n_3$ 
such that the following criteria can be applied to the line parameters $(\alpha,\gamma)$
\begin{equation}\label{eq:voigt_Accuracycriterionapplication}
\left\{\begin{array}{lll}
 \gamma/\alpha>n_2 &  \Rightarrow &  \textrm{Use the Lorentz profile for all } \nu \vspace{0.2cm} \\ 
\gamma/\alpha<n_2 & \Rightarrow  & \textrm{Use the Lorentz profile for } |\nu|>n_3\alpha  \\ 
   &  & \textrm{and the Voigt profile for } |\nu|< n_3\alpha,
\end{array}\right.
\end{equation}
which guarantees that the approximation error $|E^\mrm{V}_{\alpha,\gamma}(\nu)|<\varepsilon$ for all $\nu\in\R$.

To prove Theorem \ref{thm:Voigtaccuracy} it is convenient to normalize \eqref{eq:voigt_Ealphagammaexpr} using
the substitution $t\sqrt{\ln 2}/\alpha\leftrightarrow t$ to yield
\begin{equation}\label{eq:voigt_Ealphagammaexprnorm}
E^\mrm{V}_{\alpha,\gamma}(\nu)
=\frac{1}{\sqrt{\pi}}\int_{-\infty}^{\infty}\eu^{-t^2}\frac{t(2\tilde\nu-t)}{\tilde\gamma^2+(\tilde\nu-t)^2}\mrm{d}t
=E^\mrm{V}_{\tilde\alpha,\tilde\gamma}(\tilde\nu),
\end{equation}
and where we have introduced the dimensionless parameters
\begin{equation}\label{eq:voigt_tildepardefs}
\tilde\alpha  =  \sqrt{\ln 2}, \quad 
\tilde\gamma = \frac{\gamma}{\alpha}\sqrt{\ln 2}, \quad
\tilde\nu  =  \frac{\nu}{\alpha}\sqrt{\ln 2}. 
\end{equation}
In Appendix~\ref{sect:voigt_uniform} below is shown that the expression \eqref{eq:voigt_Ealphagammaexprnorm} converges to zero uniformly over $\tilde\nu\in\R$ 
as $\tilde\gamma\rightarrow\infty$, and hence that
\begin{equation}\label{eq:voigt_limEagres1}
\lim_{\gamma/\alpha\rightarrow\infty}\max_{\nu\in \R} |E^\mrm{V}_{\alpha,\gamma}(\nu)|=\lim_{\tilde\gamma\rightarrow\infty}\max_{\tilde\nu\in \R} |E^\mrm{V}_{\tilde\alpha,\tilde\gamma}(\tilde\nu)|=0.
\end{equation}
The result \eqref{eq:voigt_limEagres1} asserts the validity of the first statement in Theorem \ref{thm:Voigtaccuracy}.
From the error bound \eqref{eq:voigt_finalbound2} derived in Appendix~\ref{sect:voigt_uniform} it is furthermore seen that 
$|E_{\tilde\alpha,\tilde\gamma}(\tilde\nu)|$ is uniformly bounded by an arbitrary $\varepsilon>0$ for $\tilde\gamma>n_1\sqrt{\ln 2}$ 
where $n_1>0$ is fixed and the normalized frequency $|\tilde\nu|$ is sufficiently large. 
This is also consistent with the observation that $E_{\tilde\alpha,\tilde\gamma}(\tilde\nu)={\cal O}\{\tilde\nu^{-1}\}$ 
for large $\tilde\nu$, and which can be seen directly from \eqref{eq:voigt_Ealphagammaexprnorm}.
Hence, there exists a positive number $n_3$ such that 
\begin{equation}\label{eq:voigt_Eaguniformn2bound}
|E^\mrm{V}_{\tilde\alpha,\tilde\gamma}(\tilde\nu)|<\varepsilon  \hspace{2mm} \mrm{for} \hspace{2mm} |\tilde\nu|>n_3\sqrt{\ln 2},
\end{equation}
or equivalently
\begin{equation}\label{eq:voigt_limEagres2}
|E^\mrm{V}_{\alpha,\gamma}(\nu)|<\varepsilon  \hspace{2mm} \mrm{for} \hspace{2mm} |\nu|>n_3\alpha,
\end{equation}
and where $\gamma/\alpha>n_1$. The result \eqref{eq:voigt_limEagres2} finally asserts the validity of the second statement in Theorem \ref{thm:Voigtaccuracy}.

The uniform error bounds \eqref{eq:voigt_finalbound1} and \eqref{eq:voigt_finalbound2} 
derived in Appendix~\ref{sect:voigt_uniform} are useful to prove the validity of the two statements in \eqref{eq:voigt_Accuracycriterion},
but the bounds are not particularly sharp. However, since the normalized form of the relative error defined in \eqref{eq:voigt_Ealphagammaexprnorm}
can be represented by the single parameter $\tilde\gamma$, we can readily find approximate values of $n_2$ and $n_3$ yielding sharp error bounds 
by direct numerical calculation of \eqref{eq:voigt_Ealphagammadef} for various values of $\gamma/\alpha$.

\begin{figure}
\begin{center}
\includegraphics[width=0.47\textwidth]{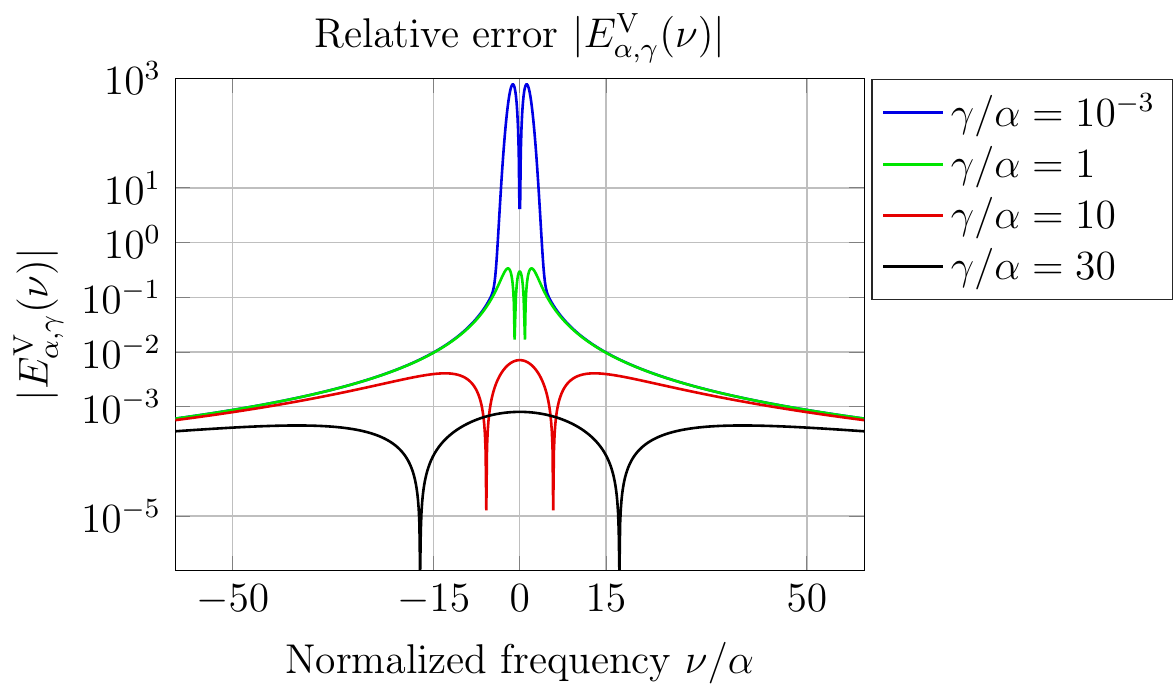}
\end{center}
\vspace{-5mm}
\caption{Computation of the relative error $|E^\mrm{V}_{\alpha,\gamma}(\nu)|$ between the Voigt profile and the Lorentz profile.
The error is plotted here as a function of the normalized frequency $\nu/\alpha$ for $\gamma/\alpha=10^{-3},1,10,30$.
}
\label{fig:voigt_matfig11.pdf}
\end{figure}

In Fig.~\ref{fig:voigt_matfig11.pdf} is shown a computation of the relative error $|E_{\alpha,\gamma}(\nu)|$ 
defined via \eqref{eq:voigt_Ealphagammadef} and \eqref{eq:voigt_tildepardefs}, 
plotted here as a function of the normalized frequency $\nu/\alpha$ for $\gamma/\alpha=10^{-3},1,10,30$.
The Voigt profile is computed based on \eqref{eq:voigt_fVjresult}  and where the 
Faddeeva function \eqref{eq:voigt_wdef} has been implemented in Matlab using the software described in \cite{Abrarov2020,Abrarov+Quine2011}.
As can be seen in this plot, there is a local maximum around $\pm1$ for small values of $\gamma/\alpha<1$,
and around $\pm\gamma/\alpha$ (in fact around $\pm 1.075\gamma/\alpha$) for large values of $\gamma/\alpha>1$. 
This means that we now have full numerical control over the localization of extremal values when executing the numerical study.
Each plot in Fig.~\ref{fig:voigt_matfig11.pdf} is made with 2000 frequency points in the interval $\nu/\alpha\in[-60,60]$, 
which is a sufficient resolution for our purposes here.

In a typical application of broadband radiative transfer in the atmosphere (see the numerical examples below), the normalized
pressure broadening is $\gamma/\alpha>10^{-3}$.
With reference to the Theorem \ref{thm:Voigtaccuracy}, we can now infer from Fig.~\ref{fig:voigt_matfig11.pdf}
that the parameter choices $n_1=10^{-3}$ together with $(n_2,n_3)=(10,15)$ and  $(n_2,n_3)=(30,50)$ 
will guarantee a relative error less than $\epsilon=10^{-2}$ and $\epsilon=10^{-3}$, respectively.

\subsection{The full Voigt profile}
Similar as above we can now define the relative error between the full Voigt and the full Lorentz profiles as
\begin{equation}\label{eq:voigt_fullEalphagammadef}
E^\mrm{FV}_{\alpha,\gamma,\nu_0}(\nu)=\frac{f_\mrm{FV}(\nu)-f_\mrm{FL}(\nu)}{f_\mrm{FL}(\nu)}, \quad \nu\in\R,
\end{equation}
and which can be manipulated to read
\begin{multline}\label{eq:voigt_fullEalphagammaexpr}
E^\mrm{FV}_{\alpha,\gamma,\nu_0}(\nu)
=\sqrt{\frac{\ln 2}{\pi}}\frac{1}{\alpha}\int_{-\infty}^{\infty}\eu^{-t^2\ln 2/\alpha^2} \\
\frac{t(2\nu-t)(\nu^4-\nu_0^4-\nu^2t(2\nu-t))}{\left((\nu_0^2-(\nu-t)^2)^2+4\gamma^2(\nu-t)^2\right)\nu^2}\mrm{d}t,
\end{multline}
which is an even function of $\nu$. The expression \eqref{eq:voigt_fullEalphagammaexpr} is obviously more involved than
\eqref{eq:voigt_Ealphagammaexpr}, but it could of course be analyzed in a similar manner as 
with the classical Voigt profile described in the Appendix \ref{sect:voigt_uniform}. 
However, this is not necessary at this point, as we can now employ more simple arguments to establish its convergence as follows.
First, it is noticed that the expression \eqref{eq:voigt_fullEalphagammaexpr} is singular at $\nu=0$,
which of course is natural since $f_{FL}(0)=0$ but $f_{FV}(0)\neq 0$. Hence, we can not have uniform convergence on the whole of $\R$. 
Secondly, we have now one more parameter to consider, the line center frequency $\nu_0$.
We therefore proceed as above, and rewrite \eqref{eq:voigt_fullEalphagammaexpr}
by using the substitution $t\sqrt{\ln 2}/\alpha\leftrightarrow t$, yielding
\begin{multline}\label{eq:voigt_fullEalphagammaexprnorm}
E^\mrm{FV}_{\alpha,\gamma,\nu_0}(\nu)
=\frac{1}{\sqrt{\pi}}\int_{-\infty}^{\infty}\eu^{-t^2} \\
\frac{t(2\tilde\nu-t)(\tilde\nu^4-\tilde\nu_0^4-\tilde\nu^2t(2\tilde\nu-t))}{\left((\tilde\nu_0^2-(\tilde\nu-t)^2)^2+4\tilde\gamma^2(\tilde\nu-t)^2\right)\tilde\nu^2}\mrm{d}t
=E^\mrm{FV}_{\tilde\alpha,\tilde\gamma,\tilde\nu_0}(\tilde\nu),
\end{multline}
where we have introduced the dimensionless parameters
\begin{equation}\label{eq:voigt_fulltildepardefs}
\tilde\alpha  =  \sqrt{\ln 2}, \quad 
\tilde\gamma = \frac{\gamma}{\alpha}\sqrt{\ln 2}, \quad
\tilde\nu  =  \frac{\nu}{\alpha}\sqrt{\ln 2}, \quad \tilde\nu_0  =  \frac{\nu_0}{\alpha}\sqrt{\ln 2}. 
\end{equation}
The scaling introduced above means that it is sufficient to consider the uniform approximation properties of \eqref{eq:voigt_fullEalphagammaexprnorm}
over the normalized frequencies $|\tilde\nu-\tilde\nu_0|< \tilde B$, and study how the approximation error behaves for a range of values of $\tilde \gamma$ 
as $\tilde\nu_0\rightarrow\infty$.
As a mathematical argument for this procedure, we can now employ the convergence of the full Lorentz profile
$f_{FL}(\nu)\rightarrow f_{L}(\nu-\nu_0)$ which is uniform for $|\nu-\nu_0|<B$ as $\nu_0\rightarrow \infty$,
\cf the derivation of \eqref{eq:voigt_uniformFL1} and \eqref{eq:voigt_uniformFL2} in Section \ref{sect:jqsrt_Lorprofiles}. 
Due to the convolution with the Gaussian profile, we immediately have also that
$f_{FV}(\nu)\rightarrow f_{V}(\nu-\nu_0)$ as well as $E^\mrm{FV}_{\alpha,\gamma,\nu_0}(\nu)\rightarrow E^\mrm{V}_{\alpha,\gamma}(\nu-\nu_0)$,
and which are uniform for $|\nu-\nu_0|<B$ as $\nu_0\rightarrow \infty$.

\begin{figure}
\begin{center}
\includegraphics[width=0.47\textwidth]{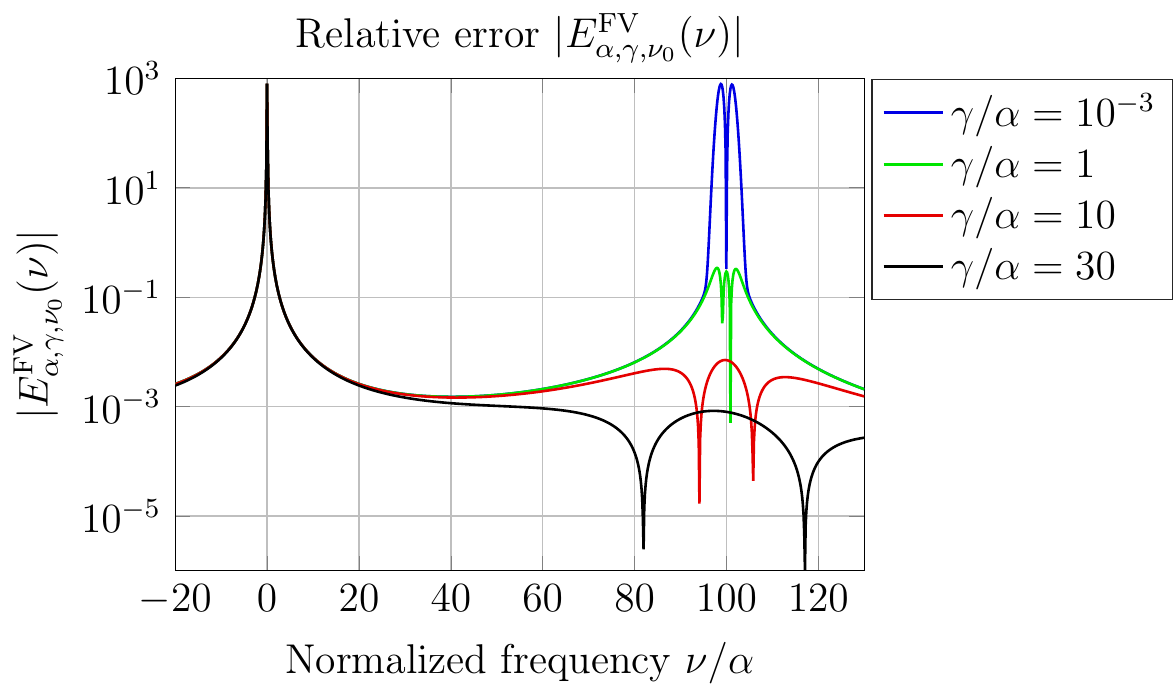}
\end{center}
\vspace{-5mm}
\caption{Computation of the relative error $|E^\mrm{FV}_{\alpha,\gamma,\nu_0}(\nu)|$ between the full Voigt profile and the full Lorentz profile.
The error is plotted here as a function of the normalized frequency $\nu/\alpha$ for $\gamma/\alpha=10^{-3},1,10,30$. 
The center frequency is $\nu_0/\alpha=100$.
}
\label{fig:voigt_matfig13.pdf}
\end{figure}

\begin{figure}
\begin{center}
\includegraphics[width=0.47\textwidth]{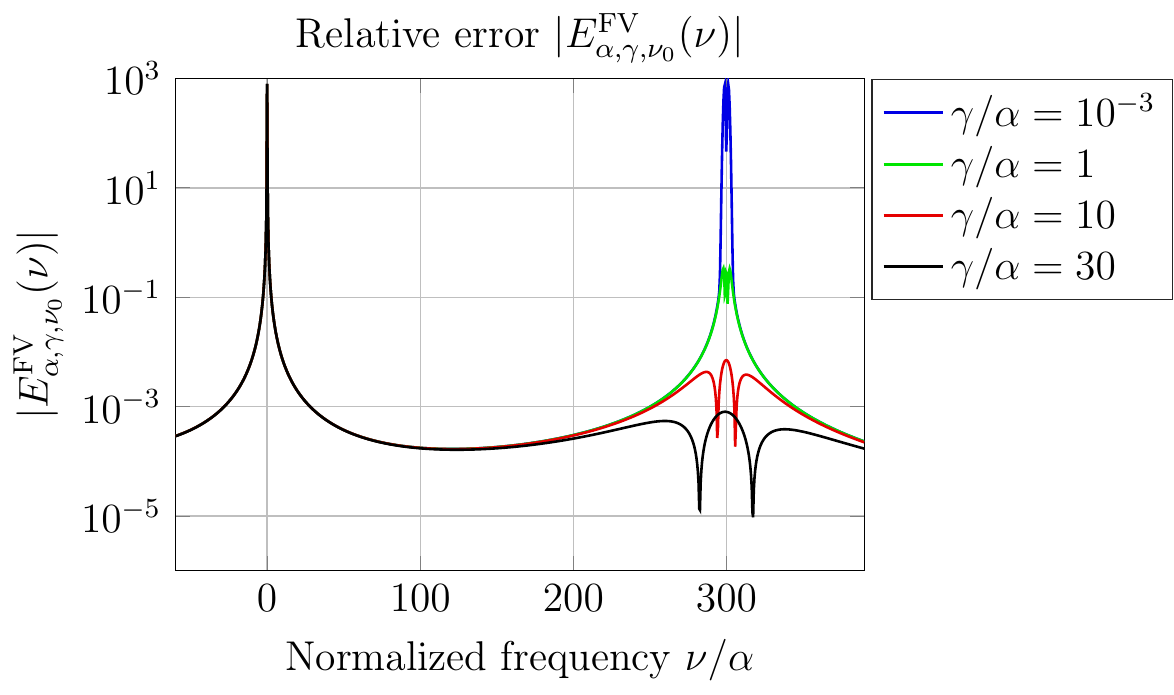}
\end{center}
\vspace{-5mm}
\caption{Same plot as in Fig.~\ref{fig:voigt_matfig13.pdf}, except here the center frequency has been increased to $\nu_0/\alpha=300$.
}
\label{fig:voigt_matfig13b.pdf}
\end{figure}

The uniform approximation procedure is illustrated in Figs.~\ref{fig:voigt_matfig13.pdf} and \ref{fig:voigt_matfig13b.pdf}.
As can be seen in these plots, the full Voigt error $E^\mrm{FV}_{\alpha,\gamma,\nu_0}(\nu)$ is locally very similar to the Voigt error
$E^\mrm{V}_{\alpha,\gamma}(\nu-\nu_0)$ of Fig.~\ref{fig:voigt_matfig11.pdf} already at $\nu_0/\alpha=100$, 
and almost identical at $\nu_0/\alpha=300$ (excluding the region close to the origin).
In a typical application of broadband radiative transfer in the atmosphere (see the numerical examples below), the normalized
center frequency is in the order of $\nu_0/\alpha>10^5$. Hence, assuming (or checking) that $\nu_0/\alpha$ is sufficiently large with reference
to the analysis illustrated in Figs.~\ref{fig:voigt_matfig13.pdf} and \ref{fig:voigt_matfig13b.pdf} above, it is safe to employ the same
Theorem \ref{thm:Voigtaccuracy}, the same approximation procedure \eqref{eq:voigt_Accuracycriterionapplication}
and the same criteria parameters $n_1$ and $(n_2,n_3)$ in connection with the full Voigt profile, as with the classical Voigt profile. 
The only difference is that the approximation is now valid over some finite region with bandwidth $B$, including the transition frequency $\nu_0$
and excluding a suitable neighborhood of the origin.

\section{Adaptive line selection}\label{sect:LS}

Even when using the fast approximation of Voigt profiles as expressed in \eqref{eq:voigt_Accuracycriterionapplication} above,
the broadband line-by-line calculations of interest in radiative transfer analysis may still be computationally huge and therefore impractical. 
Hence, it is of interest to make the computations faster and more effective by reducing the number of spectral lines that are
included in the computations and only employ the lines that are necessary and relevant at each frequency. 
There is no simple rule to make this line selection as efficient as possible in order to achieve some required accuracy.
However, based on the approximation theory that has been developed above we will outline here a simple, pragmatic and readily programmable criterion 
to make an adaptive line selection using only a few adjustable parameters to control the accuracy. 
The method is obviously suboptimal, but it is able to significantly reduce the number of spectral line calculations that are included based on simple and comprehensible criteria
while at the same time maintaining a high computational accuracy.

We consider the calculation of the absorption coefficient $k_\nu$ expressed as
\begin{equation}\label{eq:voigt_knuexpr}
k_\nu=\sum_j S_{j}f_{j}(\nu),
\end{equation}
where $S_j$ are the line strengths and $f_j(\nu)$ the approximate Voigt profiles as described in \eqref{eq:voigt_Accuracycriterionapplication}.
An individual spectral profile in \eqref{eq:voigt_knuexpr} is denoted $k_{\nu_j}(\nu)=S_jf_j(\nu)$, and which is
associated with a specific transition frequency $\nu_j$ (including pressure shift), and line parameters $\gamma_j$
and $\alpha_j$, all of which depend on height (temperature and pressure) in the atmosphere. 
It is assumed that $\gamma_j/\alpha_j>n_1$ for all $j$ and where $n_1$ is the parameter defined in Theorem \ref{thm:Voigtaccuracy}.
We will adopt here a block processing approach where the whole computational domain is divided into subdomains of relatively small and manageable sizes.
Hence, we will consider the calculation of the absorption coefficient $k_\nu$ over a fixed and relatively narrow subinterval
$\Omega=[\nu_{\mrm{a}},\nu_{\mrm{b}}]$, and estimate the contribution from all the individual spectral lines where $k_{\nu_j}(\nu)=S_jf_j(\nu)$ for $j=1,\ldots,J$.

We start by defining the minimum distance between the transition frequency $\nu_j$ and the set $\Omega$ as
\begin{equation}\label{eq:voigt_Djdef}
D_j=\left\{\begin{array}{ll}
\min\{|\nu_\mrm{a}-\nu_j|, |\nu_\mrm{b}-\nu_j|\}&  \textrm{if } \nu_j\notin\Omega, \vspace{0.2cm} \\
0 & \textrm{if } \nu_j\in\Omega, 
\end{array}\right.
\end{equation}
for all $j=1,\ldots,J$. Now, it is readily seen that for $D_j>n_3\alpha_j$ the following implication holds
\begin{multline}\label{eq:voigt_selectionineq1}
\nu\in\Omega \Rightarrow n_3\alpha_j<D_j\leq |\nu-\nu_j| \\
\Rightarrow S_j\frac{1}{\pi}\frac{\gamma_j}{\gamma_j^2+D_j^2}>
S_j\frac{1}{\pi}\frac{\gamma_j}{\gamma_j^2+|\nu-\nu_j|^2}\approx k_{\nu_j}(\nu),
\end{multline}
where $\nu_j$ is outside of $\Omega$ and where the last approximation is due to the Theorem \ref{thm:Voigtaccuracy} and 
the criterion in \eqref{eq:voigt_Accuracycriterionapplication}. 

For subintervals $\Omega$ which contain at least one spectral line $j$, we define 
the largest self-contribution over $\Omega$ as
\begin{equation}\label{eq:voigt_kmaxintdef}
k_\mrm{max}^\mrm{int}=\max_{\nu_j\in\Omega} k_{\nu_j}(\nu_j)=\max_{\nu_j\in\Omega}S_jf_j(\nu_j),
\end{equation}
which is a reasonable computational task based on the ordinary Voigt profile implemented by using \eg \cite{Abrarov2020}.
Based on \eqref{eq:voigt_selectionineq1} 
we can also estimate the largest contribution to $k_\nu$ from spectral lines $j$ outside of $\Omega$ as
\begin{equation}\label{eq:voigt_kmaxextdef}
k_\mrm{max}^\mrm{ext}=\max_{D_j>n_3\alpha_j}S_j\frac{1}{\pi}\frac{\gamma_j}{\gamma_j^2+D_j^2},
\end{equation}
and where the condition $D_j>n_3\alpha_j$ guarantees that the Lorentz approximation is valid for $\nu\in\Omega$.
An estimate of the largest contribution from any individual spectral line to $k_\nu$
is now given by $k_\mrm{max}=\max\{k_\mrm{max}^\mrm{int},k_\mrm{max}^\mrm{ext}\}$.
If $\Omega$ does not contain any spectral lines $j$, we simply choose $k_\mrm{max}=k_\mrm{max}^\mrm{ext}$.
It is now readily seen that $k_{\nu_j}(\nu)\leq k_\mrm{max}$ for $\nu\in\Omega$ and for all $j$ such that $\nu_j\in\Omega$ or $D_j>n_3\alpha_j$.

We now decide to exclude all the spectral lines with 
\begin{equation}\label{eq:voigt_selectioncritexclude}
D_j> n_3\alpha_j \quad \mrm{and} \quad S_j\frac{1}{\pi}\frac{\gamma_j}{\gamma_j^2+D_j^2}<k_\mrm{max} A,
\end{equation}
where $A$ is a small positive number. From \eqref{eq:voigt_selectionineq1}, we see that for those lines it will hold that
$k_{\nu_j}(\nu)<k_\mrm{max} A$ for all $\nu\in\Omega$. It is finally observed that the criterion in \eqref{eq:voigt_selectioncritexclude} is also equivalent to
include all the spectral lines satisfying
\begin{equation}\label{eq:voigt_selectioncritinclude}
D_j< n_3\alpha_j \quad \mrm{or} \quad S_j\frac{1}{\pi}\frac{\gamma_j}{\gamma_j^2+D_j^2}>k_\mrm{max} A.
\end{equation}
It is noted that lines satisfying $D_j< n_3\alpha_j$ are always included, \ie all lines inside or in a close neighborhood of $\Omega$.
It is also very useful to put an upper bound $K\ll J$ on the number of lines satisfying both $D_j> n_3\alpha_j$ as well as 
the second criterion in \eqref{eq:voigt_selectioncritinclude} organized in descending order.
The calculations in \eqref{eq:voigt_Djdef},  \eqref{eq:voigt_kmaxintdef}  and \eqref{eq:voigt_kmaxextdef} as well as the criterion \eqref{eq:voigt_selectioncritinclude}
can now be readily implemented in a computer code.

The extension of the procedure outlined above to the case with the full Voigt profile is straightforward. 
This means \eg that the sequence
\begin{equation}
\hat{f}_j=\frac{1}{\pi}\frac{\gamma_j}{\gamma_j^2+D_j^2}
\end{equation}
should be replaced for the sequence
\begin{equation}
\hat{f}_j=\left\{\begin{array}{ll}
\displaystyle \frac{4}{\pi}\frac{\gamma_j\nu_\mrm{b}^2}{(\nu_\mrm{a}^2-\nu_j^2)^2+4\gamma_j^2\nu_\mrm{a}^2} & \mrm{for}\ \nu_j<\nu_\mrm{a} \vspace{0.2cm} \\
\displaystyle \frac{4}{\pi}\frac{\gamma_j\nu_\mrm{b}^2}{(\nu_j^2-\nu_\mrm{b}^2)^2+4\gamma_j^2\nu_\mrm{a}^2} & \mrm{for}\ \nu_j>\nu_\mrm{b},
\end{array}\right.
\end{equation}
which will guarantee that $S_j\hat{f}_j>k_{\nu_j}(\nu)$ for $\nu\in\Omega$ and $D_j>n_3\alpha_j$, similar as in \eqref{eq:voigt_selectionineq1}.
In practice, this modification of $\hat{f}_j$ is scarcely needed in view of the very small approximation error between the Voigt and the full Voigt profiles.
The important modification here is to replace the calculation of the classical Lorentz profile with the calculation of the full Lorentz profile in the implementation 
of \eqref{eq:voigt_Accuracycriterionapplication}.

\section{Numerical examples}\label{sect:numex}
As a simple benchmark problem for comparing the accuracy and computational effort associated
with the proposed profile approximations, we consider here a broadband
line-by-line analysis of radiative transfer in the atmosphere. The computer code is implemented in Matlab and executed
on a standard laptop. 
It is emphasized that the code has not been optimized for speed. It is merely 
a straightforward implementation of a simple recursive algorithm with the aim of comparing the different
approximation methods on equal terms.

\begin{figure}
\begin{center}
\includegraphics[width=0.23\textwidth]{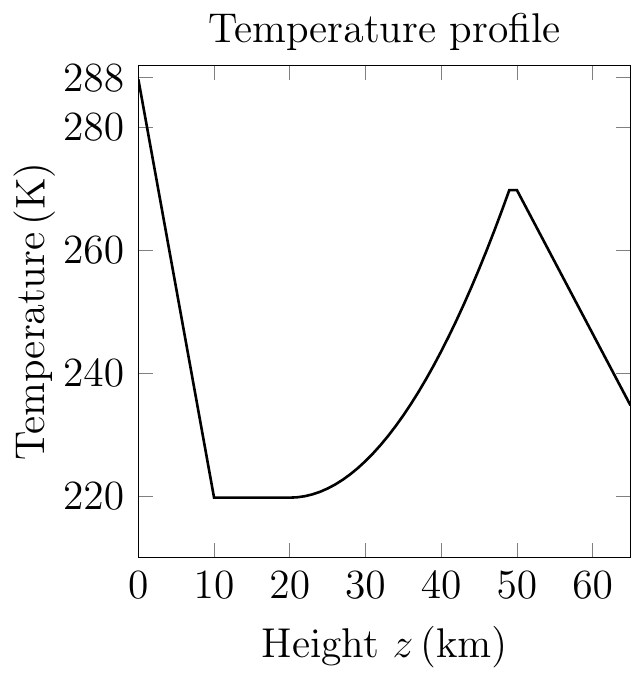}\hspace{3mm}\includegraphics[width=0.23\textwidth]{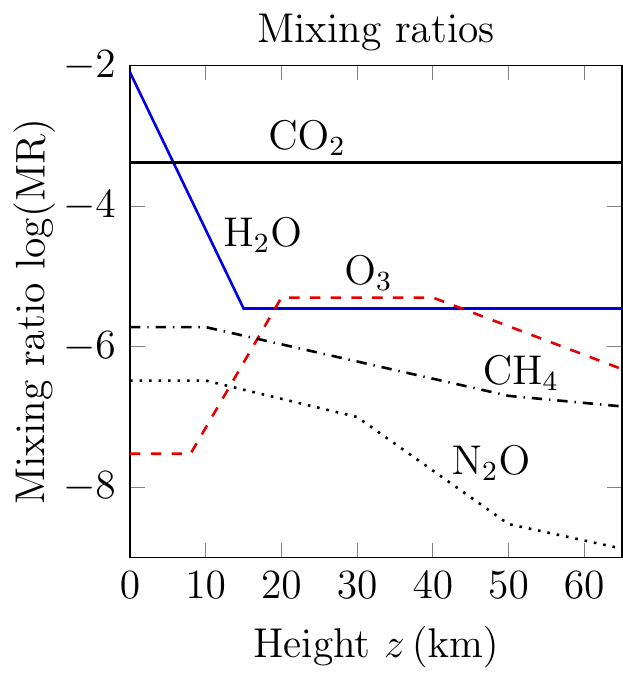}
\end{center}
\vspace{-5mm}
\caption{Representative vertical profiles of temperature and mixing ratios for the five most important greenhouse gases 
in midlatitude regions according to \cite[Figs.~3.1 and 3.2]{Liou2002}.}
\label{fig:GlobWarm_fig1o3m}
\end{figure}

We consider the computation of the outgoing monochromatic irradiance 
$F_\nu$ transmitted by the Earth at $65$\unit{km} height in a representative, plane-parallel and piece-wise homogeneous atmosphere,
as depicted in Fig.~\ref{fig:GlobWarm_fig1o3m}. 
The pressure profile is implemented as the exponential law of an isothermal atmosphere \cite[Eq. (4.24)]{Blundell+Blundell2010}
based on the mean value of the temperature profile shown in the figure.
A recursive solution of the corresponding transfer equation for thermal IR radiation in a 
non-scattering atmosphere \cite[Eq.~(4.2.2)]{Liou2002} is given by
\begin{multline}\label{eq:GlW_RTEPlaneHomsol4}
 I_\nu(z_{i+1},\mu)
= I_\nu(z_i,\mu)\eu^{-d_i\sum_s N^{(s)}(z_i)k_\nu^{(s)}(z_i)/\mu} \\
+B_\nu(T(z_i)) \left(1-\eu^{-d_i\sum_s N^{(s)}(z_i)k_\nu^{(s)}(z_i)/\mu} \right), 
\end{multline}
for $i=1,\ldots,65$. Here, $I_\nu(z_i,\mu)$ are the radiances at height $z_i$ and direction $\mu=\cos\theta$, 
$z_i=(i-1)\cdot 1$\unit{km}, $d_i=z_{i+1}-z_i$, $N^{(s)}(z_i)$ the number density of each species ($s$), $k_\nu^{(s)}(z_i)$ the 
corresponding absorption coefficients and $B_\nu(T(z_i))$ the Plack function of blackbody radiation at temperature $T(z_i)$.
The iteration is started at $z=0$ with temperature $T=288$\unit{K}.

Five different species ($s$) are included in the computation comprising a total of 430070 transitions in the range 0-3000\unit{cm^{-1}}, \cf Table~\ref{Tab:GlW_includedlines}.
The line parameters are taken from the HITRAN data base \cite{Gordon+etal2017b,Rothman+etal1998,Simeckova+etal2006}
and the following six different profile computations are considered: The Voigt (V) profile \eqref{eq:voigt_fVdef} 
and the full Voigt (FV) profile \eqref{eq:voigt_fFVdef} are based on the Faddeeva function as in \eqref{eq:voigt_fVjresult} and \eqref{eq:voigt_fFVjresult}, respectively,
and implemented by using \cite{Abrarov2020}.  
The corresponding approximations, the fast Voigt (fV) profile and the fast full Voigt (fFV) profile are based on \eqref{eq:voigt_Accuracycriterionapplication},
and their combinations with the adaptive line selection procedures (fV+LS and fFV+LS) are described in Section \ref{sect:LS}.
The parameter setting for both of the fast Voigt procedures implemented as in \eqref{eq:voigt_Accuracycriterionapplication}
are here $n_1=10^{-3}$ and $(n_2,n_3)=(10,15)$ for a maximum of 1\unit{\%} relative approximation error.
The adaptive line selection is implemented with parameters $A=10^{-8}$ and $K=1000$, see Section \ref{sect:LS}.

\begin{table}\small\centerline{
\begin{tabular}{||l|l|r||} 
\hline
{\bf Constituent} & {\bf Isotopologue}   & {\bf $\#$ lines} 0-3000\unit{cm^{-1}}   \\
\hline Water vapor   & $\mrm{H}_2^{16}\mrm{O}$ &  19615    \\
\hline Carbon dioxide  & $^{12}\mrm{C}^{16}\mrm{O}_2$  & 68829    \\
\hline Ozone  & $^{16}\mrm{O}_3$  & 210144    \\
\hline Methane  & $^{12}\mrm{CH}_4$  & 110415    \\
\hline Nitrous oxide  & $^{14}\mrm{N}_2^{16}\mrm{O}$  & 21067      \\
\hline {\bf Total $\#$ lines}  & &   {\bf 430070}      \\
\hline
\end{tabular}}
\vspace{0.1cm}
\caption{The isotopologues of the species that have been included in the modeling
and the corresponding number of transitions that are available in the HITRAN database 
over the bandwidth 0-3000\unit{cm^{-1}}.} \label{Tab:GlW_includedlines}
\end{table}

\begin{figure}
\begin{center}
\includegraphics[width=0.48\textwidth]{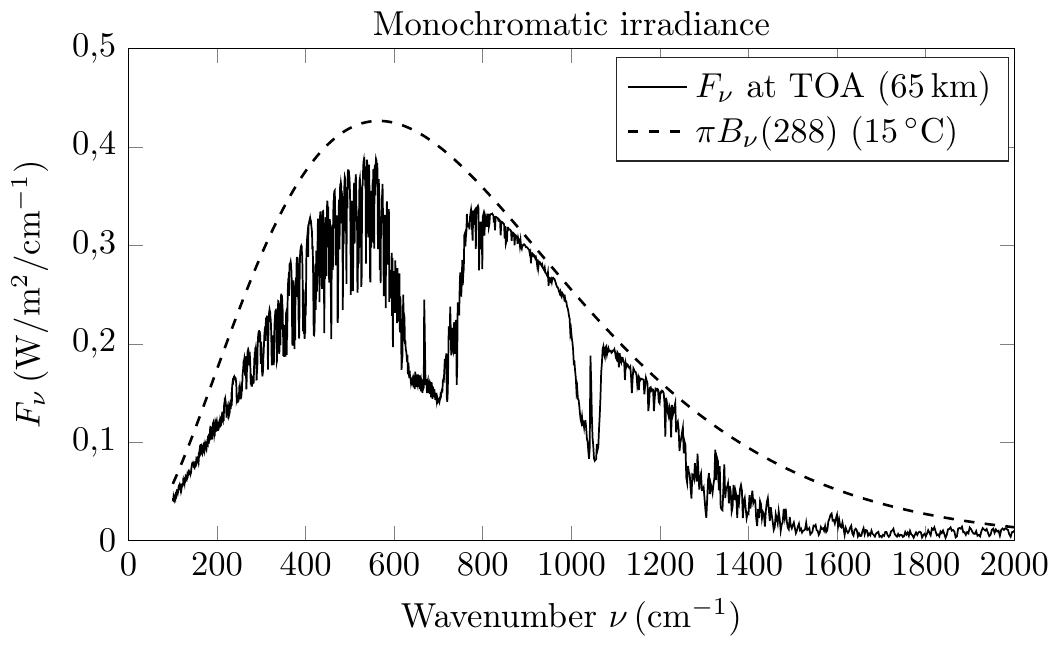}
\end{center}
\vspace{-5mm}
\caption{A computation of the outgoing monochromatic irradiance $F_\nu$ transmitted by the Earth at $65$\unit{km} height.
The computation is based on the fast Voigt approximation and the adaptive line selection procedure (fV+LS).
The dashed line shows the corresponding blackbody radiation $\pi B_\nu(T)$ at temperature $T=288$\unit{K}.}
\label{fig:GlobWarm_fig21m}
\end{figure}

The full bandwidth of the computation is 100-2000\unit{cm^{-1}} and which is divided into blocks of 2000 frequency points each.
The frequency resolution of each block is set by the Doppler broadening of the heaviest species (ozone) at the lowest temperature (220\unit{K})
yielding a total of 1955 blocks and $3.91\cdot 10^6$ frequency points over the whole bandwidth.
For each frequency block $\Omega$, the iteration in \eqref{eq:GlW_RTEPlaneHomsol4} is implemented as an array of $2000\times 10$ values of radiances
evaluated at 2000 Gauss-Legendre nodes $\nu\in\Omega$ and 10 Gauss-Legendre nodes $\mu\in(0,1)$. The integration of the total irradiance of each block
is then conveniently executed by using the corresponding Gauss-Legendre weights. The resulting monochromatic irradiance at $z=65$\unit{km} is then finally evaluated as 
the total irradiance per frequency block, as depicted in Fig.~\ref{fig:GlobWarm_fig21m}.
The frequency resolution of the monochromatic irradiance shown in the figure is hence varying from 0.15\unit{cm^{-1}} to 3.1\unit{cm^{-1}}.
Note however that the resolution of the computation in \eqref{eq:GlW_RTEPlaneHomsol4} is 2000 times more dense,
\ie the resolution varies here between 0.000075\unit{cm^{-1}} and 0.0015\unit{cm^{-1}}.
The result shown in Fig.~\ref{fig:GlobWarm_fig21m} is based on the fast Voigt profile with adaptive line selection (fV+LS) as expressed in \eqref{eq:voigt_fVjresult},
\eqref{eq:voigt_Accuracycriterionapplication} and \eqref{eq:voigt_selectioncritinclude}. The computation time on a standard laptop is about 3-4\unit{hours}.

\begin{figure}
\begin{center}
\includegraphics[width=0.47\textwidth]{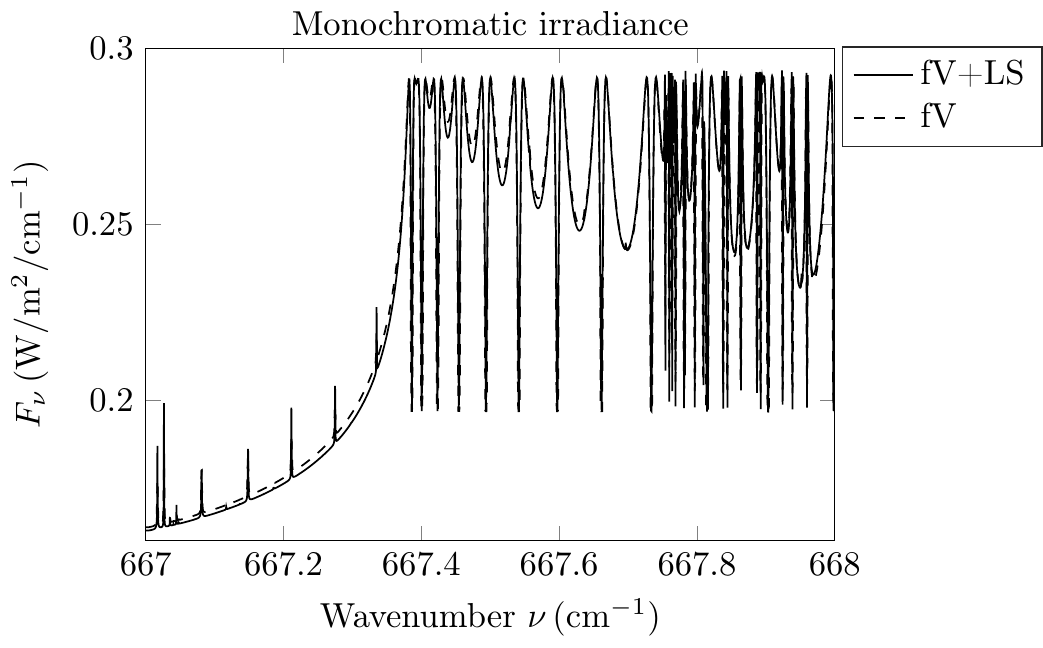}
\end{center}
\vspace{-5mm}
\caption{The outgoing monochromatic irradiance $F_\nu$ as in Fig.~\ref{fig:GlobWarm_fig21m}, evaluated here in the band
667-668\unit{cm^{-1}}. The computations are based on the fast Voigt (fV) approximation as well as the combination with
the adaptive line selection procedure (fV+LS), respectively.}
\label{fig:voigt_matfig2}
\end{figure}

\begin{figure}
\begin{center}
\includegraphics[width=0.47\textwidth]{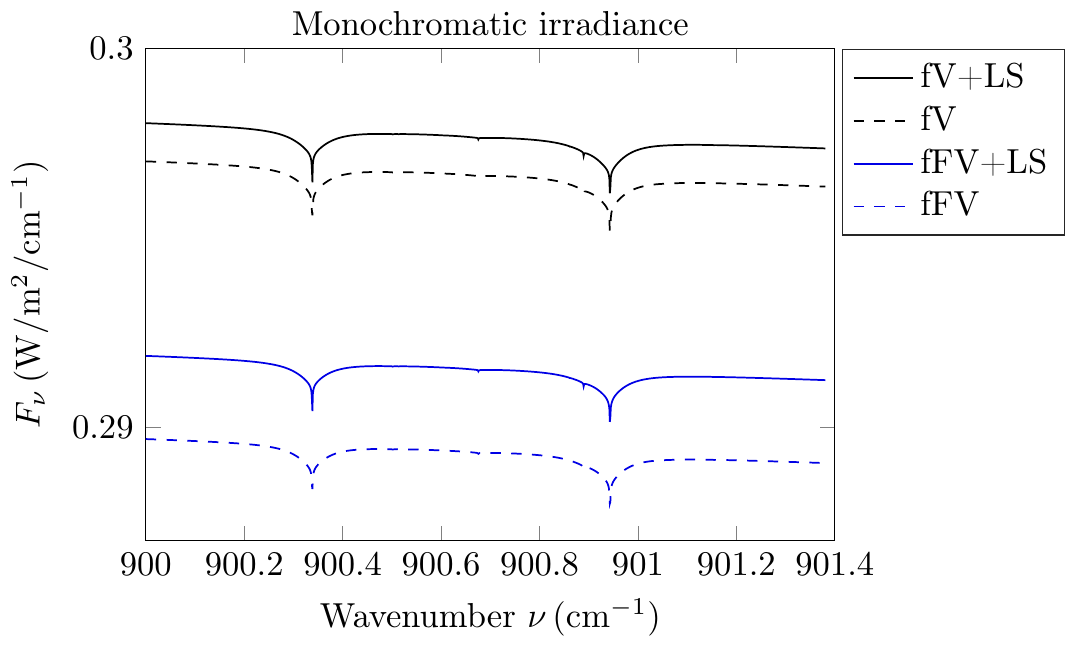}
\end{center}
\vspace{-5mm}
\caption{The outgoing monochromatic irradiance $F_\nu$ as in Fig.~\ref{fig:GlobWarm_fig21m}, evaluated here in the band
900-901.4\unit{cm^{-1}}. The computations are based on the fast Voigt (fV) as well as the fast full Voigt (fFV) approximations, and as
combinations with the adaptive line selection procedure (fV+LS, or fFV+LS), respectively.}
\label{fig:voigt_matfig3}
\end{figure}

In Figs.~\ref{fig:voigt_matfig2} and \ref{fig:voigt_matfig3} are shown the corresponding irradiance calculations for some of the profiles as mentioned above,
and which are evaluated here in the two narrow bands 667-668\unit{cm^{-1}} and 900-901.4\unit{cm^{-1}}, comprising 2000 frequency points each.
The Voigt (V) and full Voigt (FV) profiles are not shown here since they are indistinguishable from the fast approximations (fV) and (fFV) and 
which are hence very accurate in these examples.
In Fig.~\ref{fig:voigt_matfig2} we can see that there are many closely spaced absorption lines in the 667\unit{cm^{-1}} band, and there
is therefore virtually no difference between the Voigt and the full Voigt profiles, and the latter are therefore not shown in this plot. 
As we can see, there is only a small deviation between the fast Voigt (fV) and the fast Voigt with adaptive line selection (fV+LS) in this band. 

In Fig.~\ref{fig:voigt_matfig3} we can see that there are very few absorption lines in the 900\unit{cm^{-1}} band, and the far wings are therefore of more importance.
There is therefore a small difference between the Voigt and the full Voigt profiles, and there is also a small deviation using the line selection procedure.
The computer time and relative errors (relative V and FV, respectively) of these calculations are summarized in Table~\ref{Tab:jqsrt_accresults}.
The accuracy in these calculations can readily be improved by tuning the convergence parameters $(n_2,n_3)$, $A$ and $K$ described above,
and which hence can be traded against the corresponding increase in computer time.

The total number of line calculations that are available in the numerical algorithm described above is 
$430070\unit{(lines)}\cdot 65\unit{(heights)}\cdot 1955\unit{(blocks)}\approx 5.5\cdot 10^{10}$, each absorption line evaluation comprising 2000 frequency points. 
Furthermore, each of the 1955 block calculations at each of the 65 heights also consists of an evaluation 
of the radiance at the 2000 frequency points in 10 different directions, and then followed by the associated Gauss-Legendre integration.
Based on the timings presented in Table~\ref{Tab:jqsrt_accresults}, it can be estimated that a computation of the spectra as shown in Fig.~\ref{fig:GlobWarm_fig21m}
based on the Voigt profile without approximations will take about 220 days, and about twice this time using the full Voigt profile since it requires two Faddeeva evaluations instead of just one.
By using the fast Voigt (fV) approximation \eqref{eq:voigt_Accuracycriterionapplication}
the computer time for calculating all lines reduces to about 7 days. Finally, by using the adaptive line selection criteria (fV+LS) \eqref{eq:voigt_selectioncritinclude}
the number of line calculations reduces from a total of $5.5\cdot 10^{10}$ to about $4.8\cdot 10^{8}$ 
corresponding to about 0.88\unit{\%} of all the available line calculations. The final computation time on a standard laptop is now 3-4 hours.

\begin{table}\small\centerline{
\begin{tabular}{||l||l|r|c||} 
\cline{1-4} 
{\bf Band\unit{(cm^{-1})}} & {\bf Profile}   & {\bf Time\unit{(s)}} & {\bf rel. error}    \\
\hline 667-668  & fV+LS & 6  & $5.7\cdot 10^{-3}$    \\
\hline 667-668  & fV & 329 & $8.3\cdot 10^{-5}$     \\
\hline 667-668  & V & 9268 &  --   \\
\hline 667-668  & fFV+LS & 8 & $5.7\cdot 10^{-3}$      \\
\hline 667-668  & fFV & 327 & $8.3\cdot 10^{-5}$      \\
\hline 667-668  & FV & 23196 &   --    \\
\hline 
\hline  900-901.4   & fV+LS & 7 & $3.4\cdot 10^{-3}$     \\
\hline 900-901.4  & fV & 315 & $7.4\cdot 10^{-9}$      \\
\hline 900-901.4  & V & 9664 &   --   \\
\hline 900-901.4  & fFV+LS & 8 & $7.6\cdot 10^{-3}$      \\
\hline 900-901.4 & fFV & 324 & $7.2\cdot 10^{-9}$      \\
\hline 900-901.4 & FV & 23953 &   --   \\
\hline 
\end{tabular}}
\vspace{0.1cm}
\caption{Required computer time on a standard laptop 
and relative approximation errors in the computation of the total irradiance based on the 2000 frequency points 
shown in Figs.~\ref{fig:voigt_matfig2} and \ref{fig:voigt_matfig3} and the 430070 lines listed in Table \ref{Tab:GlW_includedlines}.
The reference computation corresponding to the Voigt (V) and full Voigt (FV) profiles are based on \cite{Abrarov2020}.}\label{Tab:jqsrt_accresults}
\end{table}

\section{Summary and conclusions}

This paper presents uniform error bounds for fast calculation of approximate Voigt profiles. 
The purpose is to accelerate the computationally huge broadband line-by-line analysis of radiative transfer in the atmosphere.
The main idea is to replace the relatively demanding Voigt calculations for the 
much simpler and faster Lorentz calculations whenever this can be done within a given error tolerance.
In addition, it is also demonstrated how the uniform bounds enable a fast and efficient subband adaptive line selection strategy
that includes only the spectral lines that give the most significant contribution to the absorption coefficient.
Numerical examples are included to illustrate that the two approaches can accelerate the line-by-line computations 
by several orders of magnitude with very little loss in accuracy.

A new ``full'' Voigt profile is also presented based on the ``full'' Lorentz profile, and which is obtained without making
the traditional resonance approximation. The full Lorentz profile can be very well motivated from the classical principles of molecular 
polarizabilities, optical theorems and associated sum rules connecting to the static polarizability.
On the other hand, the classical Lorentz resonance approximation can also be derived
directly from first  principles based on quantum mechanics, as in \cite[pp.~77-78]{Hartmann+etal2008}.
Hence, more research is needed to explore the potential application of this new (old!) profile.
In particular, the full Voigt profile is potentially suitable for far wing (off resonance)
calculations in spectral regions where line mixing effects can be ignored. 
Based on the numerical examples of broadband radiative transfer in the atmosphere that are presented here,
we can see that there is a slight discrepancy in the computed irradiances at high altitude
when using the Voigt and the full Voigt profiles, respectively. These discrepancies are furthermore most significant
in the so called ``infrared windows'', where the density of spectral lines is sparse and the far wing contributions are of more importance.

As a future potential of employing the approximation techniques that have been presented in this paper,
it may be of interest to explore their use to increase the accuracy and efficiency of existing line-by-line algorithms, see \eg \cite{Berk+Hawes2017}.
It may also be of interest to explore their extension to be used with existing line mixing methods  \cite{Filippov+etal2002,Filippov+Tonkov1993,Gordon1966,Gordon1967,Rosenkranz1975},
as well as with the new line shapes based on partial correlation, speed dependency and velocity changes that have
been developed recently, see \eg \cite{Ngo+etal2013,Tennyson+etal2014}. It can be expected that, while advanced methods
are employed for narrow inband calculations, it may be possible to approximate the far wing contributions by using adequate
asymptotics based on simple functions, error estimates as well as pre-calculated and stored data. 
The aim of these studies would be to render the new, more accurate line shapes and line mixing formulas more applicable for
computationally huge broad-band line-by-line analysis of radiative transfer in the atmosphere.

\appendix
\subsection{The full Voigt profile}\label{sect:voigt_FullVoigtderivation}
To show the result \eqref{eq:voigt_fFVjresult} we will employ the analytic Fourier (Laplace) transform
\begin{equation}\label{eq:voigt_Fomegadef}
F(\omega)={\cal F}\{f(t)\}=\int_{-\infty}^{\infty}f(t)\eu^{-\iu\omega t}\diff t
\end{equation}
with inverse
\begin{equation}\label{eq:voigt_ftdef}
f(t)={\cal F}^{-1}\{F(\omega)\}=\frac{1}{2\pi}\int_{-\infty+\iu y}^{\infty+\iu y}F(\omega)\eu^{\iu\omega t}\diff \omega,
\end{equation}
where $\omega=x+\iu y$ and the contour integral is carried out inside the region of analyticity of $F(\omega)$. 
We will make use of the following standard transforms
\begin{eqnarray}
{\cal F}\{\sin(at) u(t)\} & = & \frac{a}{a^2-\omega^2} \label{eq:voigt_sinatut} \\
{\cal F}\{-\iu\cos(at) u(t)\} & = & \frac{\omega}{a^2-\omega^2} \label{eq:voigt_cosatut} \\
{\cal F}\{\sqrt{\frac{b}{\pi}}\eu^{\iu\omega_0t}\eu^{-bt^2}\} & = & \eu^{-(\omega-\omega_0)^2/4b} \label{eq:voigt_embt2} 
\end{eqnarray}
where $a$ and $b$ are positive real constants, $\omega_0$ a complex valued constant and $u(t)$ the unit step (Heaviside) function. 
The region of analyticity of \eqref{eq:voigt_sinatut} and \eqref{eq:voigt_cosatut}
is $\Im\{\omega\}<0$ and for \eqref{eq:voigt_embt2} it is the whole of $\C$.
We will furthermore employ the following analytic form of the Parseval's relation
\begin{equation}\label{eq:voigt_parseval}
\int_{-\infty}^\infty f(t)g^*(t)\diff t=\frac{1}{2\pi}\int_{-\infty+\iu y}^{\infty+\iu y} F(\omega)G^*(\omega^*)\diff\omega,
\end{equation}
which can be derived directly by using \eqref{eq:voigt_ftdef} and 
where $(\cdot)^*$ denotes the complex conjugate and $y>0$ is fixed. 
The contour integral in \eqref{eq:voigt_parseval} extends along the real line ($\diff\omega=\diff x$) in the upper half of the complex plane
and requires (at least) that $F(\omega)$ is analytic in a neighborhood of the line $\omega=+\iu y$ in the upper half plane and
that $G(\omega)$ is analytic in a neighborhood of the line $\omega=-\iu y$ in the lower half.

In order to derive \eqref{eq:voigt_fFVjresult}, we start from \eqref{eq:voigt_fFVdef} and write
\begin{multline}
f_\mrm{FV}(\nu)=f_\mrm{G}(\nu)*f_\mrm{FL}(\nu) \\
=\sqrt{\frac{\ln 2}{\pi}}\frac{1}{\alpha}\eu^{-\nu^2\ln 2/\alpha^2} * \Im\left\{ \frac{2}{\pi}\frac{\nu}{\nu_0^2-\nu^2-\iu 2\gamma\nu} \right\}
\end{multline}
where $*$ denotes convolution and where \eqref{eq:voigt_hnudef}, \eqref{eq:voigt_FLdef} and \eqref{eq:voigt_fGdef} have been used.
This means that we can write $f_\mrm{FV}(\nu)=\Im\left\{ h_\mrm{FV}(\nu)\right\}$ where
\begin{equation}
h_\mrm{FV}(\nu)=\sqrt{\frac{\ln 2}{\pi}}\frac{2}{\pi\alpha}\int_{-\infty}^\infty \eu^{-(\nu-x)^2\ln 2/\alpha^2}\frac{x\diff x}{\nu_0^2-x^2-\iu 2\gamma x}.
\end{equation}
By introducing $\omega=x+\iu\gamma$, $\omega_0=\nu+\iu\gamma$, $a^2=\nu_0^2-\gamma^2$ 
and $b=\alpha^2/4\ln 2$, the integral above can be written
\begin{multline}
I=\int_{-\infty}^\infty \eu^{-(\nu-x)^2\ln 2/\alpha^2}\frac{x\diff x}{\nu_0^2-x^2-\iu 2\gamma x} \\
=\int_{-\infty+\iu\gamma}^{\infty+\iu\gamma}\eu^{-(\omega-\omega_0)^2\ln 2/\alpha^2}\frac{\omega-\iu\gamma}{a^2-\omega^2}\diff\omega \\
=\int_{-\infty+\iu\gamma}^{\infty+\iu\gamma}\eu^{-(\omega-\omega_0)^2/4b}\frac{\omega}{a^2-\omega^2}\diff\omega \\
-\frac{\iu\gamma}{a}\int_{-\infty+\iu\gamma}^{\infty+\iu\gamma}\eu^{-(\omega-\omega_0)^2/4b}\frac{a}{a^2-\omega^2}\diff\omega,
\end{multline}
and where it is assumed that $\nu_0>\gamma>0$ and hence $a>0$. 
By using the Parseval's relation \eqref{eq:voigt_parseval} and the standard transforms \eqref{eq:voigt_sinatut} through \eqref{eq:voigt_embt2}, it now follows that
\begin{multline}
I=2\pi\int_0^\infty\sqrt{\frac{b}{\pi}}\eu^{\iu\omega_0t}\eu^{-bt^2}(-\iu)^*\cos(at)\diff t \\
-\frac{\iu\gamma}{a}2\pi \int_0^\infty\sqrt{\frac{b}{\pi}}\eu^{\iu\omega_0t}\eu^{-bt^2}\sin(at)\diff t.
\end{multline}
By rewriting the trigonometric functions using exponentials the integral above can also be written in the more generic form
\begin{multline}
I=\frac{\sqrt{\pi}\alpha}{2\sqrt{\ln 2}} 
\left( \iu\int_0^\infty\eu^{-bt^2+\iu(\omega_0+a)t}\diff t \right. \\
+\iu\int_0^\infty\eu^{-bt^2+\iu(\omega_0-a)t}\diff t  
 -\frac{\gamma}{a}\int_0^\infty\eu^{-bt^2+\iu(\omega_0+a)t}\diff t \\
\left. +\frac{\gamma}{a}\int_0^\infty\eu^{-bt^2+\iu(\omega_0-a)t}\diff t \right).
\end{multline}
All the integrals above are now in a form where we can substitute and complete the squares inside the exponent to get
\begin{multline}
\int_0^\infty \eu^{-\alpha^2t^2/4\ln 2-ct}\diff t \\
=\frac{2\sqrt{\ln 2}}{\alpha}\eu^{c^2\ln 2/\alpha^2}\int_{0}^\infty\eu^{-(t+c\sqrt{\ln 2}/\alpha)^2}\diff t \\
=\frac{2\sqrt{\ln 2}}{\alpha}\eu^{c^2\ln 2/\alpha^2}\int_{c\sqrt{\ln 2}/\alpha}^\infty\eu^{-t^2}\diff t \\
=\frac{2\sqrt{\ln 2}}{\alpha}\eu^{c^2\ln 2/\alpha^2}\frac{\sqrt{\pi}}{2}\mrm{erfc}(c\sqrt{\ln 2}/\alpha) \\
=\frac{\sqrt{\pi\ln 2}}{\alpha}w(\iu c\sqrt{\ln 2}/\alpha),
\end{multline}
where $c=-\iu(\omega_0\pm a)$, $\mrm{erfc}(z)$ is the complementary error function 
and $w(z)=\eu^{-z^2}\mrm{erfc}(-\iu z)$ the Faddeeva function, see \cite[Eq.~7.2.1--7.2.3]{Olver+etal2010}.
Collecting all the results above, we get 
finally
\begin{multline}\label{eq:voigt_fFVjresult2}
h_\mrm{FV}(\nu)=\sqrt{\frac{\ln 2}{\pi}}\frac{1}{\alpha}\left(
\left(-\frac{\gamma}{a}+\iu \right)w\left((\nu+a+\iu\gamma)\frac{\sqrt{\ln 2}}{\alpha}\right) \right. \\
\left. +\left(\frac{\gamma}{a}+\iu \right)w\left((\nu-a+\iu\gamma)\frac{\sqrt{\ln 2}}{\alpha}\right)
\right).
\end{multline}

\subsection{Uniform convergence of the Voigt profile}\label{sect:voigt_uniform}
We consider the uniform converge of the error term
\begin{equation}\label{eq:voigt_Ealphagammaexprnorm2}
E^\mrm{V}_{\tilde\alpha,\tilde\gamma}(\tilde\nu)
=\frac{1}{\sqrt{\pi}}\int_{-\infty}^{\infty}\eu^{-t^2}\frac{t(2\tilde\nu-t)}{\tilde\gamma^2+(\tilde\nu-t)^2}\mrm{d}t,
\end{equation}
where $\tilde\alpha=\sqrt{\ln 2}$, $\tilde\gamma=(\gamma/\alpha)\sqrt{\ln 2}$ and $\tilde\nu=(\nu/\alpha)\sqrt{\ln 2}$ as defined in \eqref{eq:voigt_Ealphagammaexprnorm}.
Assume that $\tilde\nu>0$, let $0<a<1$ be a suitable chosen parameter and make the following
estimates of $|\tilde\nu-t|$ in the respective subintervals
\begin{equation}
\left\{\begin{array}{lll}
t\leq 0 & \Rightarrow & |\tilde\nu-t| \geq \tilde\nu \hspace{0.2cm} \\
0\leq t \leq a\tilde\nu & \Rightarrow & |\tilde\nu-t| \geq (1-a)\tilde\nu \hspace{0.2cm} \\
a\tilde\nu \leq t \leq 2\tilde\nu & \Rightarrow & |\tilde\nu-t| \geq 0 \hspace{0.2cm} \\
t\geq 2\tilde\nu & \Rightarrow & |\tilde\nu-t| \geq \tilde\nu.
\end{array}\right.
\end{equation}
An error estimate based on \eqref{eq:voigt_Ealphagammaexprnorm2} is now readily obtained as 
\begin{multline}\label{eq:voigt_Errorestimate1}
|E^\mrm{V}_{\tilde\alpha,\tilde\gamma}(\tilde\nu)|
\leq \frac{1}{\sqrt{\pi}}\int_{-\infty}^{0}\eu^{-t^2}\frac{(-t)(2\tilde\nu-t)}{\tilde\gamma^2+\tilde\nu^2}\mrm{d}t \\
+ \frac{1}{\sqrt{\pi}}\int_{0}^{a\tilde\nu}\eu^{-t^2}\frac{t(2\tilde\nu-t)}{\tilde\gamma^2+(1-a)^2\tilde\nu^2}\mrm{d}t \\
+ \frac{1}{\sqrt{\pi}}\int_{a\tilde\nu}^{2\tilde\nu}\eu^{-t^2}\frac{t(2\tilde\nu-t)}{\tilde\gamma^2}\mrm{d}t \\
+ \frac{1}{\sqrt{\pi}}\int_{2\tilde\nu}^{\infty}\eu^{-t^2}\frac{t(t-2\tilde\nu)}{\tilde\gamma^2+\tilde\nu^2}\mrm{d}t.
\end{multline}
By making the substitution $t\leftrightarrow -t$ in the first integral, and then repeating the whole procedure for negative $\tilde\nu$, it is found that
\begin{multline}\label{eq:voigt_Errorestimate2}
|E^\mrm{V}_{\tilde\alpha,\tilde\gamma}(\tilde\nu)|
\leq \frac{1}{\sqrt{\pi}}\int_{0}^{\infty}\eu^{-t^2}\frac{t(2|\tilde\nu|+t)}{\tilde\gamma^2+\tilde\nu^2}\mrm{d}t \\
+ \frac{1}{\sqrt{\pi}}\int_{0}^{a|\tilde\nu|}\eu^{-t^2}\frac{t(2|\tilde\nu|-t)}{\tilde\gamma^2+(1-a)^2\tilde\nu^2}\mrm{d}t \\
+ \frac{1}{\sqrt{\pi}}\int_{a|\tilde\nu|}^{2|\tilde\nu|}\eu^{-t^2}\frac{t(2|\tilde\nu|-t)}{\tilde\gamma^2}\mrm{d}t \\
+ \frac{1}{\sqrt{\pi}}\int_{2|\tilde\nu|}^{\infty}\eu^{-t^2}\frac{t(t-2|\tilde\nu|)}{\tilde\gamma^2+\tilde\nu^2}\mrm{d}t.
\end{multline}
The right hand side of \eqref{eq:voigt_Errorestimate2} can now be evaluated by using the following integrals
\begin{equation}
\frac{1}{\sqrt{\pi}}\int_{R_1}^{R_2}\eu^{-t^2}t\mrm{d}t=\frac{1}{2\sqrt{\pi}}\left(\eu^{-R_1^2}-\eu^{-R_2^2}\right),
\end{equation}
and
\begin{multline}
\frac{1}{\sqrt{\pi}}\int_{R_1}^{R_2}\eu^{-t^2}t^2\mrm{d}t \\
=\frac{1}{2\sqrt{\pi}}\left(R_1\eu^{-R_1^2}-R_2\eu^{-R_2^2}\right)+\frac{1}{4}\left( \mrm{erf}(R_2)-\mrm{erf}(R_1)\right),
\end{multline}
and where $\mrm{erf}(\cdot)$ is the error function \cite[Eq.~7.2.1]{Olver+etal2010}. By carrying out the integrations above
the following upper bound is obtained
\begin{multline}\label{eq:voigt_finalbound1}
|E^\mrm{V}_{\tilde\alpha,\tilde\gamma}(\tilde\nu)|\leq
\frac{1}{\tilde\gamma^2+\tilde\nu^2}\left(\frac{1}{2}+\frac{|\tilde\nu|}{\sqrt{\pi}}-\frac{1}{4}\mrm{erf}(2|\tilde\nu|) \right) \\
+\frac{1}{\tilde\gamma^2+(1-a)^2\tilde\nu^2}\left(\frac{2|\tilde\nu|}{2\sqrt{\pi}}(1-\eu^{-a^2\tilde\nu^2}) \right. \\
\left. + \frac{a|\tilde\nu|}{2\sqrt{\pi}}\eu^{-a^2\tilde\nu^2} -\frac{1}{4}\mrm{erf}(a|\tilde\nu|)\right) \\
+\frac{1}{\tilde\gamma^2}\left( \frac{(2-a)|\tilde\nu|}{2\sqrt{\pi}}\eu^{-a^2\tilde\nu^2} -\frac{1}{4}(\mrm{erf}(2|\tilde\nu|)-\mrm{erf}(a|\tilde\nu|))\right).
\end{multline}
Finally, the bound in \eqref{eq:voigt_finalbound1} can be further relaxed and simplified as
\begin{multline}\label{eq:voigt_finalbound2}
|E^\mrm{V}_{\tilde\alpha,\tilde\gamma}(\tilde\nu)|\leq
\frac{1}{\tilde\gamma^2+\tilde\nu^2}\left(\frac{1}{2}+\frac{|\tilde\nu|}{\sqrt{\pi}} \right) \\
+\frac{1}{\tilde\gamma^2+(1-a)^2\tilde\nu^2}\frac{(2+a)|\tilde\nu|}{2\sqrt{\pi}} 
+\frac{1}{\tilde\gamma^2} \frac{(2-a)|\tilde\nu|}{2\sqrt{\pi}}\eu^{-a^2\tilde\nu^2}.
\end{multline}
It can readily be seen that the bound in \eqref{eq:voigt_finalbound2} converges to zero uniformly over $\tilde\nu\in\R$ as $\tilde\gamma\rightarrow \infty$.
It can furthermore be seen that  \eqref{eq:voigt_finalbound2} is uniformly bounded by an arbitrarily small number $\varepsilon>0$ 
for $\tilde\gamma>n_1\sqrt{\ln 2}$ where $n_1>0$ is given and $|\tilde\nu|>n_3\sqrt{\ln 2}$ where $n_3$ is a sufficiently large positive real number.


\end{document}